\documentclass[CMYK,NoSecthm]{jcomsec_author}  %author version
\usepackage{savesym}
\savesymbol{AND}
\usepackage[linkcolor=black,citecolor=black,colorlinks=false,bookmarksnumbered]{hyperref} %Printed Version
\usepackage{multicol}
\usepackage{enumerate}
\usepackage{amssymb}
\usepackage{setspace}
\usepackage{mathtools}
\usepackage{colortbl}
\usepackage{color}
\usepackage{ctable}
\usepackage{wrapfig}
\usepackage[numbers,sort&compress]{natbib}
\usepackage{hypernat}
\usepackage[all]{hypcap}
\usepackage{array}
\usepackage{microtype}
\usepackage{doi}

\DisableLigatures[f]{encoding = *, family = * }

\newcolumntype{X}[1]{>{\centering\arraybackslash}m{#1}}

\usepackage{amsmath,tabulary,graphicx,times,caption,fancyhdr}
\usepackage[utf8]{inputenc}
\usepackage{ifpdf}
\usepackage{algorithmic}
\usepackage{pifont}
\usepackage[nointegrals]{wasysym}
\usepackage{hhline}
\usepackage{subcaption}
\usepackage{color}
\begin{document}

\begin{frontmatter}

\title{{\color{red} Performance Evaluation of Apache Spark MLlib Algorithms on CSE-CIC-IDS2018 Intrusion Detection Dataset}}

\author[DLS]{Ramin Atefinia}
\ead{ramina@post.com \rm(R. Atefinia)}
\author[DLS,CorAuth]{Mahmood Ahmadi}
\ead{m.ahmadi@razi.ac.ir \rm(M. Ahmadi)}
\address[DLS]{Department of Computer Engineering and Information Technology, Razi University, Iran.}

%\address{\ignorespaces Department of Computer Engineering and Information Technology\unskip, 
 % \institution{Razi University}\unskip, Kermanshah\unskip, Kermanshah\unskip, \cny{Iran}
 %}

%\let\runningauthors\runningauthor
%\runningtitle{Performance Evaluation of Intrusion Detection Systems}%%\runningauthors{Atefinia and Ahmadi }
\corauth[CorAuth]{Mahmood Ahmadi.}


%\twocolumn[
%\begin{@twocolumnfalse}
%\maketitle
\begin{abstract}
{\color{red} The rise of internet and web services} usage and the emergence of the fifth generation of cellular network technology (5G) along with ever-growing Internet of Things (IoT) data traffic will grow global internet usage. To ensure the security of future networks, Machine Learning based Intrusion Detection and Prevention Systems (IDPS) should be implemented to detect novel attacks, and big data parallel processing tools can be used to handle an extensive collection of training data in these systems. In this paper Apache Spark, a general-purpose and fast cluster computing platform is used for processing and training a large volume of network traffic feature data. In this work, {\color{red}the most important features} of the CSE-CIC-IDS2018 dataset are used for the construction of machine learning models, and then the most popular machine learning approaches, namely Logistic Regression, Support Vector Machine (SVM), three different Decision Tree Classifiers and Naive Bayes algorithm are used to train the model using up to eight number of worker nodes. Our Spark cluster contains seven machines acting as worker nodes, and one machine is configured as master and worker. We use the CSE-CIC-IDS2018 dataset to evaluate the overall performance of these algorithms on Botnet attacks, and distributed hyperparameter tuning is used to find the best single decision tree parameters. Experimental results indicate that Apache Spark can significantly enhance the training time of intrusion detection models when using more worker nodes in algorithms such as naive Bayes and random forest and handle the problem {\color{red}of high memory} usage in algorithms like MLlib SVM and decision trees with high depth, while the accuracy of the trained models {\color{red}remains unchanged. We have tried to depict results of six algorithms and determine the best parallelizable algorithm for anomaly-based intrusion detection system training.} Using an extensive amount of worker nodes adds overhead to our application and {\color{red} more speed-up} can not be achieved in some algorithms. Spark-based hyperparameter tuning can be used to find the best suitable hyperparameters instead of the widely-used {\color{red} manual trial and error method.} We have achieved up to 100\% accuracy using selected features via the learner method in our experiments.
\end{abstract}
 \begin{keyword}
 Intrusion detection systems,  Apache spark, MLlib, Machine learning
 \end{keyword}
\end{frontmatter} 

% \end{@twocolumnfalse}
% ]

\section{Introduction}
The ever-increasing network traffic data and the growth of network-based services require new data processing techniques and tools. The risks of an insecure network in our era are so costly that every business must protect their IT infrastructure with high priority. The development and progress of the internet of things devices and the implementation of the fifth generation of cellular network technology (5G) will offer prime targets for intruders \cite{andreev2014internet}. Machine learning-based or anomaly-based intrusion detection is a broad field of research that attempts to detect new and zero-day intrusions using machine learning algorithms. The limitations of {\color{red}the current intrusion detection systems} that use signature-based methods are listed below.
\begin{itemize}
\item Only the known attacks can be detected and a small change in attack pattern can bypass the signature check.
\item To process all signatures by IDS, high usage of memory and computational resources is needed. 
\item Noise in traffic can significantly increase the false-alarm rate.
\item Outdated signature databases leave IDS completely vulnerable to new attacks.
\end{itemize}

The machine learning-based IDS systems are currently under research, and the challenges include higher false alarm rates and building a comprehensive profile of normal behaviors. \\
In the recent few years, the use of big data solutions like Apache Spark and Apache Hadoop to process network data on a large scale is on the rise because standards like Message Passing Interface (MPI) are at the wrong level of abstraction. Spark, on the other hand, is also automatically fault-tolerant, and by utilizing various libraries like MLlib, the researchers do not need to reinvent and re-architect their entire needs.\\
The MLlib library provides APIs that can combine multiple machine learning algorithms into one workflow \cite{meng2016mllib}. An implemented model can read a column from a DataFrame, map it to a feature vector and produce a new DataFrame with an extra column of appended predicted labels. The scaling capability of each algorithm in MLlib depends on the size of the dataset and the available memory on each cluster nodes, However, the problem of distributing a memory-intensive task over a large number of worker nodes is possible for all of the supported algorithms.\\
Model selection and tuning of hyperparameters is another capability of a Spark cluster. Using CrossValidator and TrainValidationSplit tools in Spark, we can tune an algorithm or pipeline and specify a metric for the fitted model. An example of a tuned hyperparameter could be the stopping criteria of each algorithm. This process may take a huge amount of time, but the results can be used to efficiently train a model in the future. \\
Six MLlib algorithms are used in this work. Logistic regression is a  generalized linear model and can be used to predict a binary outcome among Botnet and Benign flow instances. This algorithm can also be used for a multiclass dataset using the multinomial logistic regression method. LinearSVC in Spark MLlib supports binary classification tasks with linear Support Vector Machine.\\
Two types of naive Bayes algorithms (multinomial and Bernoulli) are supported in MLlib. The training data in this algorithm is used only once and no caching is required. Thus the speed-up of this method in spark is very high. Three families of decision tree classification, namely single decision tree, random forest, and Gradient-boosted tree (GBT) classifier are also implemented in this work. \\

The main contributions of this paper are as follows: In this paper, first, the dataset will be preprocessed before working with machine learning algorithms. Then we use the learner method to drop less useful features in the Botnet attacks of the CSE-CIC-IDS2018 dataset. Distributed hyperparameter tuning method is proposed instead of {\color{red} manual trial and error} to find the best parameters for a single decision tree algorithm, and then the speedup of Apache Spark parallel processing using six MLlib algorithms is depicted for a different type of worker nodes, and accuracy metrics of our models are calculated.\par 

The rest of this paper is organized as follows. In section \ref{relatedwork}, we review publications that attempt to use big data tools for intrusion detection or prevention. In section \ref{ids} Apache Spark and its capabilities for intrusion detection are discussed. In section \ref{dataset}, the details of our dataset are discussed. In section \ref{results} our experimental results are presented, and finally, the conclusion  is presented in section \ref{conclusion}.
\section{Related Work}
\label{relatedwork}
In recent years a few works are done using big data frameworks to detect intrusions. In \cite{belouch2018performance}, they used four classification algorithms and measured the performance of naive Bayes, Decision Tree, Random Forest, Support Vector Machine using Apache Spark and UNSW-NB15 dataset with all 42 features. They used the MLlib library for experiments and the final analysis of the results. The configuration of Spark and the number of nodes used are not mentioned in the paper; thus the speed-up of the Spark framework is not clearly understood by this work.\par

{\color{red}In \cite{dobson2018performance} both KDD99 and NSL-KDD datasets are used to measure performance metrics of MLlib algorithms such as Logistic Regression, Random Forest, Naïve Bayes, SVM, GB Trees and MLP in a virtual machine environment. The experiments are performed for binary and multiclass scenarios but the speedup of Apache Spark is not presented in this work.}
\par

 In \cite{gupta2016framework}, they used Apache Spark and along with NSL-KDD and KDD Cup 1999 intrusion detection datasets \cite{protic2018review}. They used two feature selection algorithms, namely, Chi-squared and correlation-based feature selection. They also used five common classification-based intrusion detection techniques: Support Vector Machine, Logistic Regression, Gradient Boosted Decision trees, Random Forest, and Naive Bayes. The configuration of Spark and the effect of the number of executors on training time are not calculated in this work.\par
In \cite{hsieh2016detection}, they proposed a DDoS attack detection technique based on Artificial Neural Networks; The proposed technique is implemented on the Apache Spark cluster. They used the 2000 DARPA LLDOS 1.0 dataset to apply their experiments. The resulting detection systems can detect attacks in real-time with over 94\% detection rates.\par

In \cite{kumar2017performance}, they performed distributed Sentiment Analysis on Apache Spark and Message Passing Interface (MPI). They evaluated the performance and overhead time associated with the computing time on spark using 100GB, 500GB, and 1TB datasets, respectively. The results show that CPU utilization is high in the MPI processing framework, but MPI is approximately 2x times faster than Spark processing over a cluster of 10 machines with 40 cores. This work shows that the memory utilization factor in Spark programming can be controlled by the Spark job controller. In contrast, the memory utilization factor could be adjusted in MPI programming for better performance. The High-Performance Computing Library, MPI thus, can be utilized in high-performance and big data applications to improve speed-ups.\par
In \cite{terzi2017big}, public data was inspected with a novel unsupervised anomaly detection technique on a cluster of Apache Spark in Azure HDInsight. They used the CTU-13 dataset for botnet traffic analysis. The final results were visualized as Three-dimensional by dimension reduction technique known as Principal component analysis (PCA).\par
In \cite{dahiya2018network}, they used Apache Spark for network intrusion detection in datasets with big size. They used Canonical Correlation Analysis (CCA) and Linear Discriminant Analysis (LDA) for feature selection in the UNSW-NB 15 dataset, and then the performance of seven classification algorithms is compared. The Apache Spark speed-up is not calculated in this work.
\par
In \cite{marir2018distributed}, they proposed a distributed deep learning approach to detect abnormal behavior based on big data analytics and the multi-layer ensemble learning technique. The utilized model is a distributed deep belief network as a dimensionality reduction method, and then an iterative reduce method for multi-layer ensemble SVM based on Spark is performed. The developed system shows high performance in the detection of abnormal situations in a distributed manner.\par
{\color{red}In \cite{saravanan2020performance} NSL-KDD dataset is used to test the performance metrics for Logistic Regression, Decision Tree and SVM using Apache Spark and up to 96.8\% accuracy achieved for Decision Tree algorithm. The exact configuration of Apache Spark is not mentioned in this work. }
\par
In this work, we use Spark's machine learning capabilities to train six {\color{red} parallelizable algorithms.} To do this, first, we configure Spark and then create a Spark context. The training data is loaded to a dataset, and hyperparameters required by algorithms are determined. The trained model is used to predict the type of attack in test data, and then performance measures are calculated. Finally, the trained models can be saved to local files for practical and future use. {\color{red} In this work, we try to answer the following research questions: 1. Which algorithm can be better parallelized in an intrusion detection system training scenario, and 2. What is the effect of the number of spark nodes on performance measures?} A method for the selection of the best hyperparameters in a distributed manner is also presented.

\section{Intrusion Detection using Apache Spark}
\label{ids}
In this section, the Apache spark framework and its MLlib library are presented and then the machine learning algorithm for Intrusion Detection is described.
\subsection{Apache Spark Framework}
Apache Spark is parallel processing, and cluster computing framework that extends the well-known MapReduce model to support various models of computations such as stream processing and interactive queries \cite{karau2015learning}. Spark is more efficient than MapReduce and can run computations in memory. Spark offers APIs in Python, Java, Scala, SQL, and various libraries. Spark is a computational engine at its core and distributes computational tasks across many worker nodes\cite{zaharia2016apache}. In Spark, it is possible to write a machine learning application to classify data in real-time.\par
\begin{figure*}[!h]
\centering
\includegraphics[width=0.9\textwidth]{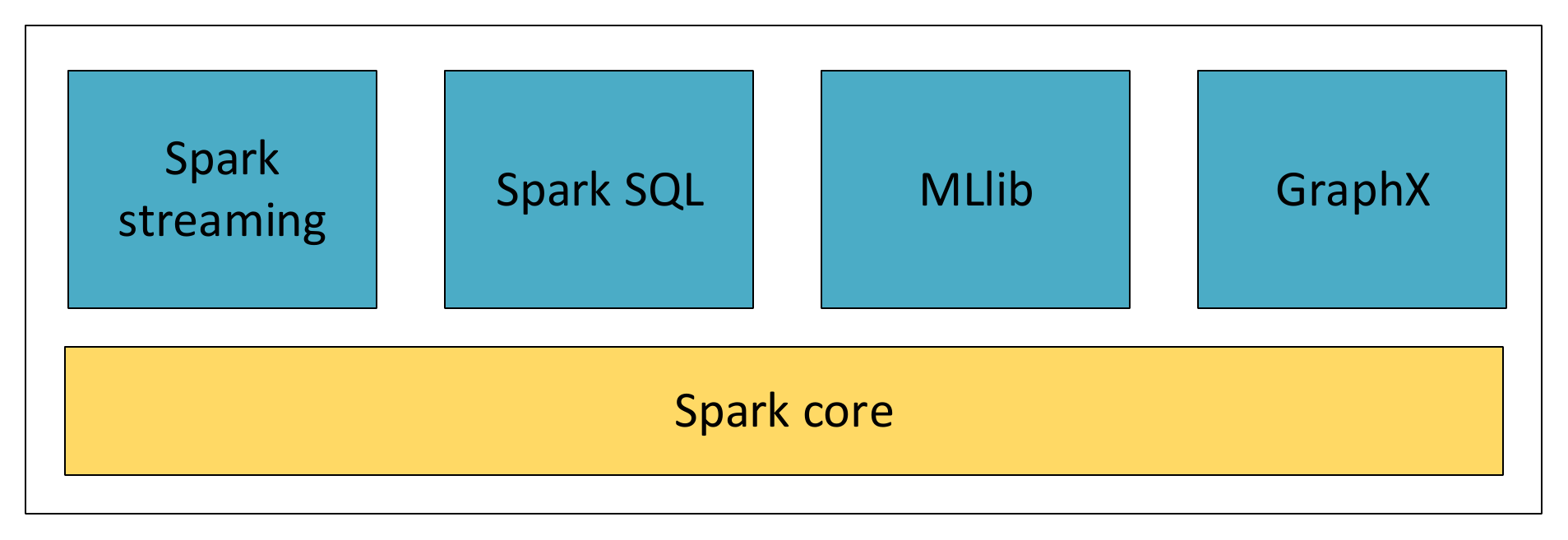}
\caption{Spark Components, Including MLlib, GraphX, Spark SQL, Spark Streaming on top of Spark Core}
\label{fig-SparkComponents}
\end{figure*}

 Spark is not an altered version of Hadoop and has no dependency on it. Spark comes with its cluster manager and uses Hadoop for storage purposes only. As shown in Figure \ref{fig-SparkComponents}, MLlib, GraphX, Spark SQL, Spark Streaming, and Spark Core are the main elements of Spark framework.\par

Spark Core is the base component of the Apache Spark project. It provides task scheduling, task dispatching, and basic I/O functionalities in a distributed manner. Spark core functionalities are exposed through an API for Java, Python, R, .NET, and Scala programming languages.

Spark Streaming provides high-throughput, fault-tolerant, and scalable stream processing of data streams. The data streams can be received from sources like Kafka or TCP sockets and can be processed using Spark's high-level functions. It is possible to apply machine learning and graph processing algorithms to various data streams.

In Spark SQL, a data abstraction named DataFrame is introduced, and it is suitable for processing structured data. Spark SQL interface provides more information about the structure of data and computations compared to RDDs. It is possible to interact with this component using either SQL or Dataset API.

GraphX is an API in Apache Spark, used for graph and graph-parallel computations. Examples of graph computations include disaster detection systems, page rank, business analysis, fraud detection, and geographic information systems. Spark comes with a growing number of graph algorithms to simplify the mentioned graph tasks.

\subsection{Resilient Distributed Dataset (RDD)}
Spark's main primary data structure and abstraction for working with provided data are called resilient distributed datasets or RDD. An RDD is an unchanging distributed collection of elements or objects. In RDD, each dataset is divided into logical partitions. The logical partitions can be computed on several worker nodes with several numbers of executors. RDD can be created by loading our dataset, and then transformations and actions can be operated on this data\cite{karau2017high}.\par
\begin{figure*}[!htb]
\centering
\includegraphics[width=0.9\textwidth]{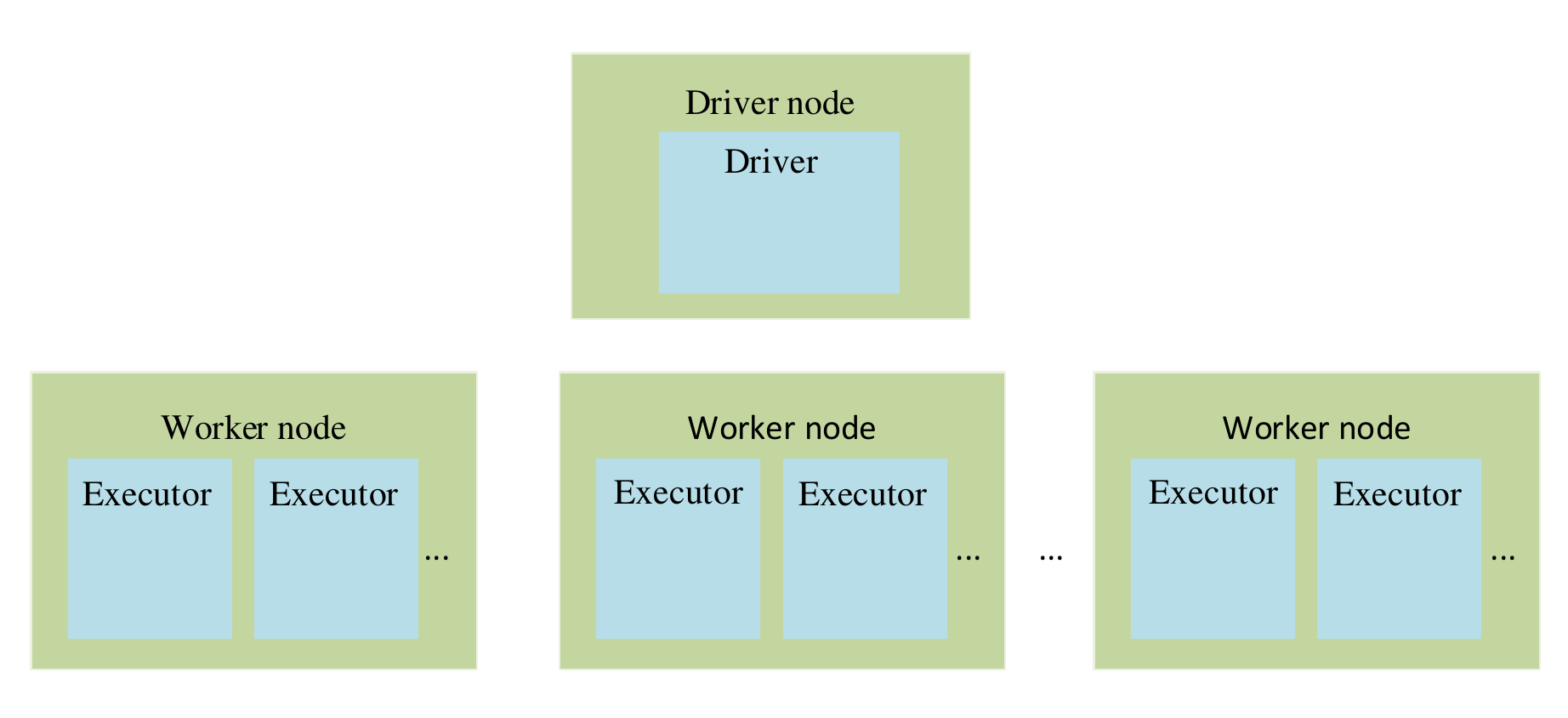}
\caption{The driver program creates multiple workers, which read data from distributed file system.}
\label{fig-Spark-runtime}
\end{figure*}

Spark exposes RDDs through an API, where datasets are represented as objects, and transformations are done using methods on these objects. It is possible to call a persist method to mark RDDs for future reuse in operations. Spark keeps these datasets in memory, but in case of low RAM, RDDs can be spilled to disk. Figure \ref{fig-Spark-runtime} shows Spark runtime. As depicted in this Figure, the driver program creates multiple workers, which read data from the distributed file system. The computed RDD partitions can be persisted in memory\cite{zaharia2012resilient}. 
\subsection{Spark Dataframe and Dataset APIs}
Unlike Resilient Distributed Datasets, in a Spark DataFrame, data is organized like a  relational database and into named columns. This is similar to relational databases like MySQL. DataFrames can be constructed from existing RDDs or other sources. The API for DataFrame is available in python, java, Scala, and R programming languages, but it is represented differently.\par
A dataset, on the other hand, is a distributed collection of data. In addition to the advantages of an RDD, a dataset has the benefits of an optimized execution engine in Spark SQL. The API for Dataset is available in Java and Scala.

\subsection{MLlib and Spark parallelizable algorithms}
In this section, several machine learning techniques of MLlib are presented.
\subsubsection{MLlib, Scalable Machine Learning on Spark}
Spark is equipped with a distributed machine learning library named MLlib that benefits from data parallelism to store and perform operations on data. Algorithms like principal component analysis for dimensionality reduction are also included in this library. We can use the Spark yarn cluster to train a model using MLlib in Scala, Java, and python. Spark standalone cluster supports MLlib programs using Java and Scala programming languages. RScripts are not tested in this work. In the following subsections, we discuss these parallelizable algorithms and implement them on our datasets.

\subsubsection{Logistic regression}
Logistic regression is a very efficient method for two-class classification tasks. The logistic function or sigmoid function is used at the core of this method. This function can take a real number and at the output a value between 0 and 1 is produced:

\iftrue
\begin{equation}
\label{SEF}
f(x) = \frac{L}{1 + e^{-k(x-x_0)}}
\end{equation}
\fi

In Eq. \ref{SEF}, $e$ is Euler’s number, and $x_0$ is the $x$ value of the sigmoid's middle point of the line segment. $L$ is the maximum value of the curve, and $k$ is the steepness of the curve or growth rate.
If we model out network traffic to normal and Bot attack, then the probability for being an attack or regular network usage is calculated \cite{wright1995logistic}.  \par
We used Spark DataFrame to load our dataset, and the CSV format is used instead of LIBSVM. The features are then transformed into FeatureVector. A feature column is then added to the Spark DataFrame using VectorAssembler feature transformer in Spark. Our dataset contains two classes, and a label column can be added to DataFrame \cite{clasreg}. The high memory usage problem in this algorithm is solved by a distributed Spark cluster.  \par

\subsubsection{Support Vector Machine}
A Support Vector Machine (SVM) is a supervised discriminative classifier that can also be used for regression. Each data item in SVM can be represented as a point in an n-dimensional space, where n is representing the number of features in the dataset. The value of a particular coordinate is the value of the corresponding features in the dataset. The task of classification is done by finding a hyperplane that can properly separate two classes \cite{amarappa2014data}. \par
The basic idea of a support vector machine like a neural net is an optimal hyperplane for linearly separable patterns. The linear SVM is a standard technique for large-scale classification problems.

\subsubsection{Naive Bayes}
Naive Bayes is a simple classification algorithm based on Bayes’ Theorem. Using Bayes’ Theorem, it is possible to find the probability of an event occurring given the probability of another event that has occurred before. The following mathematical equation depicts Bayes’ theorem:

\iftrue
\begin{equation}
\label{bayes}
P(A|B)=\frac{P(B|A)P(A)}{P(B)}
\end{equation}
\fi

Assuming y is the class variable and a dependent feature vector of size n is shown by X, Bayes’ theorem can be applied to a dataset as bellow:

\iftrue
\begin{equation}
\label{bayesapplied}
P(y|X)=\frac{P(X|y)P(y)}{P(X)}
\end{equation}
\fi

\subsubsection{Decision Trees}
Decision trees are a set of prevalent supervised classification algorithms. A decision tree is like a flowchart made of nodes and branches. At each of the nodes, data is split based on one of the features, which in result generates two or more branches. This continues until further creation of branches is no longer possible.
Apache Spark MLlib library supports decision trees for multiclass and binary classification and for regression, using both categorical and continuous features. The algorithm partitions data by rows. This makes possible the distributed training with millions of instances\cite{treesMLlib}.\par
Random Forest Tree and Gradient Boosted Tree are ensemble methods using decision trees as base learners. Random Forest creates a large number of decision trees based on bootstrap aggregating or bagging. It resamples the data and for each sample trains a new classifier. Gradient Boosted Tree adds a classifier at a time so that the next classifier is trained to improve the already trained ensemble.

\paragraph{Information gain and Node impurity}
The node impurity shows how well the decision tree splits the data, and it shows the homogeneity of the labels at the node. MLlib provides two options for impurity measures in classification tasks (entropy and Gini impurity) and one for regression (variance). Table \ref{NodeImpurity} 
describes node impurity in MLlib.

\begin{table}[!htb]
\caption{Node impurity and information gain.}
\label{NodeImpurity}
\setlength{\tabcolsep}{12pt}
\renewcommand{\arraystretch}{2}
\begin{tabular}{|l|l|l|}
\hline
\textbf{Impurity} & \textbf{Task}  & \textbf{Formula}                        \\ \hline
Gini impurity     & Classification & $\sum_{i=1}^{C}f_i(1-f_i)$              \\ \hline
Entropy           & Classification & $\sum_{i=1}^{C}-f_i\log(f_i)$            \\ \hline
Variance          & Regression     & $\frac{1}{N}\sum_{i=1}^{N}(y_i-\mu )^2$ \\ \hline
\end{tabular}
\end{table}

In Table \ref{NodeImpurity}, $C$ is the number of unique labels, and $f_i$ is the frequency of label $i$ at a node. $y_i$ is label for an instance, N represents the number of instances and $\mu$ is the mean given by $\frac{1}{N} \sum_{i=1}^{N}y_i$.

\subsubsection{Learner Based Feature Selection}
The variable selection, also known as attribute selection or feature selection is the process of choosing the most important features of a given dataset. In a network intrusion detection dataset, there might be several features that do not contribute to the detection of intrusion. So in order to reduce overfitting, improve the accuracy of the model, and reduce the training time, we can carry out feature selection before training the model. In this paper, we use an analyzer function to evaluate the performance of algorithms with different subsets of the dataset to find the best one that results in better accuracy. The selected features then will be used to train our six machine learning algorithms. Table \ref{tab2} depicts highly relevant features selected from our datasets.

\begin{table*}[!htb]
\centering
  \caption{Description of the selected features of CSE-CIC-IDS2018 dataset}
\begin{tabular}{ll}
\rowcolor[HTML]{C0C0C0} 
Feature Name     & Description                                                  \\
Dst Port         & Destination port                                             \\
\rowcolor[HTML]{EFEFEF} 
Fwd Pkt Len Mean & Mean value of packet in forward direction                  \\
Flow IAT Min     & Minimum time among two flows                               \\
\rowcolor[HTML]{EFEFEF} 
Fwd IAT Tot      & Total time among two packets sent in the forward direction
\end{tabular}
\label{tab2}
\end{table*}

\section{Datasets Specification}
\label{dataset}
In this section, the CSE-CIC-IDS2018 dataset and distributed hyperparameter tuning method in Spark are presented.
\subsection{CSE-CIC-IDS2018 dataset}
The CSE-CIC-IDS2018 dataset contains realistic and modern types of attacks and benign (normal) records. According to the available features\cite{sharafaldin2018toward}\cite{tavallaee2009detailed}\cite{sharafaldin2018toward}, the Botnet attack subset is suitable to compare the performance of MLlib algorithms. This subset contains 1048575 samples after removing the outliers. The pattern of traffic in infected Internet-connected devices can be recognized by most of these algorithms. To load this dataset into a Spark dataset, it needs to be converted to LIBSVM or proper CSV format. Table \ref{attacks} shows a list of executed attacks in a simulated network in AWS and the duration of each executed attack in the CSE-CIC-IDS2018 dataset. The obtained results are validated using a 3-fold cross-validation of the model.

\begin{table*}[!h]
\centering
  \caption{ Executed attacks list, tools and duration}
\label{attacks}
\begin{tabular}{lllll}
\cline{1-1}
\rowcolor[HTML]{C0C0C0} 
\multicolumn{1}{|c|}{\cellcolor[HTML]{C0C0C0}\textbf{Attack}} & \multicolumn{1}{c}{\cellcolor[HTML]{C0C0C0}\textbf{Tools}}                                                                                        & \multicolumn{1}{c}{\cellcolor[HTML]{C0C0C0}\textbf{Duration}} & \multicolumn{1}{c}{\cellcolor[HTML]{C0C0C0}\textbf{Attacker}} & \multicolumn{1}{c}{\cellcolor[HTML]{C0C0C0}\textbf{Victim}}                                   \\ \cline{1-1}
Bruteforce attack                                             & \begin{tabular}[c]{@{}l@{}}FTP – Patator\\ SSH – Patator\end{tabular}                                                                             & One day                                                       & Kali linux                                                    & Ubuntu 16.4 (Web Server)                                                                      \\
\rowcolor[HTML]{EFEFEF} 
DoS attack                                                    & \begin{tabular}[c]{@{}l@{}}Hulk, GoldenEye,\\ Slowloris, Slowhttptest\end{tabular}                                                                & One day                                                       & Kali linux                                                    & Ubuntu 16.4 (Apache)                                                                          \\
DoS attack                                                    & Heartleech                                                                                                                                        & One day                                                       & Kali linux                                                    & Ubuntu 12.04 (Open SSL)                                                                       \\
\rowcolor[HTML]{EFEFEF} 
Web attack                                                    & \begin{tabular}[c]{@{}l@{}}Damn Vulnerable Web App (DVWA)\\ In-house selenium framework\\ (XSS and Brute-force)\end{tabular}                      & Two days                                                      & Kali linux                                                    & Ubuntu 16.4 (Web Server)                                                                      \\
Infiltration attack                                           & \begin{tabular}[c]{@{}l@{}}First level: Dropbox download in a\\ windows machine.\\ Second Level: Nmap and portscan\end{tabular}                   & Two days                                                      & Kali linux                                                    & Windows Vista and Macintosh                                                                   \\
\rowcolor[HTML]{EFEFEF} 
Botnet attack                                                 & \begin{tabular}[c]{@{}l@{}}Ares (developed by Python): remote shell,\\ file upload/download, capturing\\ screenshots and key logging\end{tabular} & One day                                                       & Kali linux                                                    & \begin{tabular}[c]{@{}l@{}}Windows Vista, 7, 8.1,\\  10 (32-bit) and 10 (64-bit)\end{tabular} \\
DDoS+PortScan                                                 & \begin{tabular}[c]{@{}l@{}}Low Orbit Ion Canon (LOIC) for UDP,\\ TCP, or HTTP requests\end{tabular}                                               & Two days                                                      & Kali linux                                                    & \begin{tabular}[c]{@{}l@{}}Windows Vista, 7, 8.1,\\  10 (32-bit) and 10 (64-bit)\end{tabular}
\end{tabular}
\end{table*}

Table \ref{features} shows example features of the CSE-CIC-IDS2018 dataset.

\begin{table*}[!h]
\centering
\caption{Description of example features extracted from network traffic in CSE-CIC-IDS2018 dataset}
\label{features}
\begin{tabular}{ll}
\rowcolor[HTML]{C0C0C0} 
\multicolumn{1}{c}{\cellcolor[HTML]{C0C0C0}\textbf{Feature Name}} & \multicolumn{1}{c}{\cellcolor[HTML]{C0C0C0}\textbf{Description}}   \\
fl\_dur                                                           & Flow duration                                                      \\
\rowcolor[HTML]{EFEFEF} 
tot\_fw\_pk                                                       & Total packets in the forward direction                             \\
tot\_bw\_pk                                                       & Total packets in the backward direction                            \\
\rowcolor[HTML]{EFEFEF} 
tot\_l\_fw\_pkt                                                   & Total size of packet in forward direction                          \\
fw\_pkt\_l\_max                                                   & Maximum size of packet in forward direction                        \\
\rowcolor[HTML]{EFEFEF} 
fw\_pkt\_l\_min                                                   & Minimum size of packet in forward direction                        \\
fw\_pkt\_l\_avg                                                   & Average size of packet in forward direction                        \\
\rowcolor[HTML]{EFEFEF} 
fw\_pkt\_l\_std                                                   & Standard deviation size of packet in forward direction             \\
Bw\_pkt\_l\_max                                                   & Maximum size of packet in backward direction                       \\
\rowcolor[HTML]{EFEFEF} 
Bw\_pkt\_l\_min                                                   & Minimum size of packet in backward direction                       \\
Bw\_pkt\_l\_avg                                                   & Mean size of packet in backward direction                          \\
\rowcolor[HTML]{EFEFEF} 
Bw\_pkt\_l\_std                                                   & Standard deviation size of packet in backward direction            \\
fl\_byt\_s                                                        & flow byte rate that is number of packets transferred per second    \\
\rowcolor[HTML]{EFEFEF} 
fl\_pkt\_s                                                        & flow packets rate that is number of packets transferred per second \\
fl\_iat\_avg                                                      & Average time between two flows                                    
\end{tabular}
\end{table*}

\subsubsection{Distributed Hyperparameter Tuning Method in Spark}
Hyperparameter tuning involves finding the best set of parameters to give to our algorithm to achieve the best accuracy measures. Generally, there are two techniques that are used for this purpose, namely grid search and random search. In the grid search method, every possible list of values with every combination is evaluated, and in the random search method, random combinations of parameters are tested to find the best possible values for the model. In this work, we utilized the random search method to find the best hyperparameters in single decision tree experiments. In a Random Forest algorithm, parameters such as the number of trees and the depth of the tree can be examined using Apache Spark. Then the combination found with the highest accuracy can be used to further perform our experiments. For example, we can observe that using a depth of seven for a single decision tree can achieve the best performance in terms of training time and accuracy. Therefore, in future experiments, we can only use a depth of seven as one of our parameters to reduce the training time with no impact on accuracy. We achieved high true positive rates and low false-positive rates using this method in a single decision tree algorithm.

\section{Experimental Results}
\label{results}
In the following section, the experimental results of deploying machine learning algorithms with the intrusion detection dataset using the Apache Spark framework are presented. We use a Spark cluster of one master and worker node and seven worker nodes to demonstrate the performance of the apache spark framework. After preprocessing the datasets, to speed up the training phase, first, we fit our analyzer function to training sets only and when the chosen components can be used to speed up training time. The results are mapped to both the training set and the test set. The experiments are performed using eight cluster nodes, each with 2 virtual CPU cores of the Intel Xeon E3-12xx v2 family and 4GB of RAM, and one executor for each node. We have used Spark's own standalone cluster manager in these experiments, and the latency between the master and each worker node is less than one millisecond.

In each experiment, the performance of the algorithm is calculated using TP, TN, FP, FN. Where P = Positive, N = Negative, T = True and F = false \cite{heberlein2007statistical}. True Positive or TP is the number of correctly classified attack records. True Negative or TN is the number of correctly classified normal records. False Positive or FP is the number of incorrectly classified normal records. False Negative or FN is the number of incorrectly classified attack records. Using the following equations, these performance measures can be calculated:\\

%\noindent
True Positive Rate (Recall): \\
\begin{equation}
\label{TPR}
TPR=\frac{TP}{(FN+TP)}
\end{equation}

True Negative Rate: \\
\begin{equation}
\label{TNR}
TNR=\frac{TN}{(TN+FP)}
\end{equation}

False Positive Rate:
\begin{equation}
\label{FPR}
FPR=\frac{FP}{(FP+TN)}
\end{equation}

False Negative Rate:
\begin{equation}
\label{FNR}
FNR={FN}{(FN+TP)}
\end{equation}

Accuracy: \\
\begin{equation}
\label{ACC}
ACC=\frac{(TN + TP)}{(TN+TP+FN+FP)} 
\end{equation}

F1 score:\\
\begin{equation}
\label{F1}
F1=\frac{(2TP)}{(2TP+FP+FN)} 
\end{equation}

\subsection{Implementation}
\begin{center}
\begin{figure}[!h]
\centering
\includegraphics[scale=0.7]{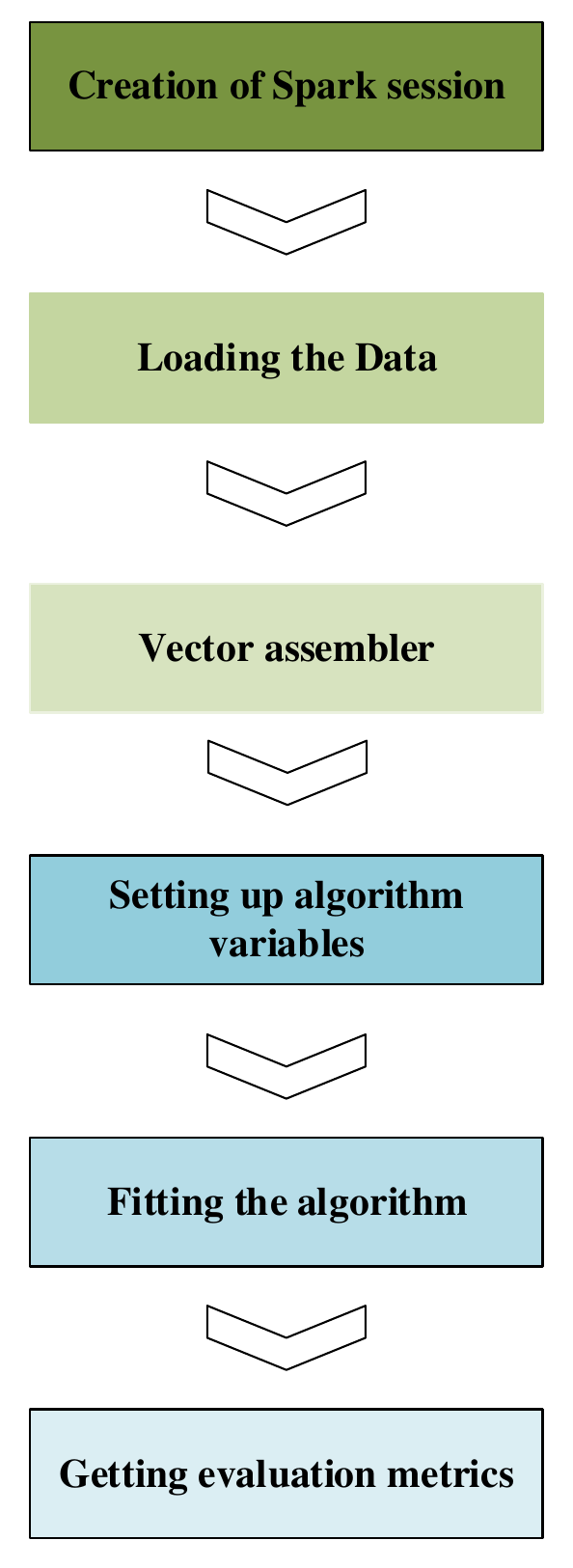}
\caption{Implementation steps}
\label{imp}
\end{figure}
\end{center}

{\color{red} The implementation steps are depicted in Figure 
\ref{imp}. Every Spark application requires a SparkSession. To perform} experiments, we first load the dataset containing headers with the inferSchema option. We need two columns for the representation of data in Spark: Features and Labels. By using VectorAssembler class, we create our features array. Then the fitting step will be performed on the selected algorithm. Figure \ref{imp} depicts the implementation steps of our method.\\
A single decision tree experiment is performed using the obtained hyperparameters in the tuning step. A very deep tree could potentially allow a higher accuracy in some applications, but the higher cost of training and overfitting may occur. In our application, a depth of seven can achieve maximum performance, and more than 32 bins will increase training time in a single decision tree algorithm, however, the number of correctly classified records can not outperform our previous work that utilized a modular neural network model\cite{modular} but the results are still comparable to a deep neural network model in \cite{basnet2019towards}.

\subsection{Performance Measures}
Tables \ref{table1} - \ref{table8} depicts performance measures for Botnet attacks in the CSE-CIC-IDS2018 dataset by using one to eight executors. In some algorithms like Naive Bayes, we can observe more speed-ups compared to other algorithms. The communication overhead, data communication cost, and unparallelizable operations may result in lower performance in some algorithms by increasing the number of worker nodes; however, the problem of error in memory is eliminated by using a distributed cluster. According to some past studies, our eight-node cluster setup suffices to compare the scalability of the tested algorithms \cite{bekkerman2011scaling}. \\

\begin{table*}[!h]
\caption{Performance measures for Botnet attacks in CSE-CIC-IDS2018 dataset with one active worker}
\resizebox{1\textwidth}{!}{
\label{table1}
\begin{tabular}{|l|c|c|c|c|c|c|c|c|}
\hline
\rowcolor[HTML]{C0C0C0} 
\multicolumn{1}{|c|}{\cellcolor[HTML]{C0C0C0}Algorithm} & TN     & FP    & FN     & TP     & Precision & Recall & F-Measure & Training Time (Seconds) \\ \hline
Logistic Regression                                     & 691043 & 71341 & 148246 & 137945 & 0.779     & 0.791  & 0.779     & 4653.09                 \\ \hline
\rowcolor[HTML]{EFEFEF} 
Support Vector Machine                                  & 715711 & 46673 & 286189 & 2      & 0.519     & 0.683  & 0.590     & 9550.03                 \\ \hline
Bernoulli naive Bayes                                   & 762304 & 80    & 441    & 285750 & 1.000     & 1.000  & 1.000     & 195.02                  \\ \hline
\rowcolor[HTML]{EFEFEF} 
Single Decision Tree (Tuned)                            & 762345 & 39    & 116    & 286075 & 1.000     & 1.000  & 1.000     & 643.86                  \\ \hline
Random Forest Tree                                      & 738865 & 23519 & 9581   & 276610 & 0.969     & 0.968  & 0.969     & 338.01                  \\ \hline
\rowcolor[HTML]{EFEFEF} 
Gradient Boosted Tree                                   & 762304 & 80    & 4419   & 281772 & 0.996     & 0.996  & 0.996     & 4673.48                 \\ \hline
\end{tabular}
}
\end{table*}

\begin{table*}[!h]
\caption{Performance measures for Botnet attacks in CSE-CIC-IDS2018 dataset with two active workers}
\resizebox{1\textwidth}{!}{
\label{table2}
\begin{tabular}{|l|c|c|c|c|c|c|c|c|}
\hline
\rowcolor[HTML]{C0C0C0} 
\multicolumn{1}{|c|}{\cellcolor[HTML]{C0C0C0}Algorithm} & TN     & FP    & FN     & TP     & Precision & Recall & F-Measure & Training Time (Seconds) \\ \hline
Logistic Regression                                     & 691043 & 71341 & 148246 & 137945 & 0.779      & 0.791  & 0.779     & 3057.47                 \\ \hline
\rowcolor[HTML]{EFEFEF} 
Support Vector Machine                                  & 715711 & 46673 & 286189 & 2     & 0.519     & 0.683  & 0.590     & 6071.14                \\ \hline
Bernoulli naive Bayes                                   & 762304 & 80    & 441    & 285750 & 1.000     & 1.000  & 1.000     & 90.75                  \\ \hline
\rowcolor[HTML]{EFEFEF} 
Single Decision Tree (Tuned)                            & 762351 & 33    & 112    & 286079 & 1.000     & 1.000  & 1.000     & 572.07                  \\ \hline
Random Forest Tree                                      & 743958 & 18426 & 5254   & 280937 & 0.978     & 0.977  & 0.978     & 317.82                  \\ \hline
\rowcolor[HTML]{EFEFEF} 
Gradient Boosted Tree                                   & 762305 & 79    & 4419   & 281772 & 0.996     & 0.996  & 0.996     & 3295.91                 \\ \hline
\end{tabular}
}
\end{table*}

\begin{table*}[!htb]
\caption{Performance measures for Botnet attacks in CSE-CIC-IDS2018 dataset with three active workers}
\label{table3}
\resizebox{1\textwidth}{!}{
\begin{tabular}{|l|c|c|c|c|c|c|c|c|}
\hline
\rowcolor[HTML]{C0C0C0} 
\multicolumn{1}{|c|}{\cellcolor[HTML]{C0C0C0}Algorithm} & TN     & FP    & FN     & TP     & Precision & Recall & F-Measure & Training Time (Seconds) \\ \hline
Logistic Regression                                     & 691043 & 71341 & 148246 & 137945 & 0.779     & 0.791  & 0.779     & 2436.68                 \\ \hline
\rowcolor[HTML]{EFEFEF} 
Support Vector Machine                                  & 715711 & 46673 &  286189  & 2      & 0.519     & 0.683  & 0.590     & 3600.08                 \\ \hline
Bernoulli naive Bayes                                   & 762304 & 80    & 441    & 285750 & 1.000     & 1.000  & 1.000     & 71.55                  \\ \hline
\rowcolor[HTML]{EFEFEF} 
Single Decision Tree (Tuned)                            & 762341 & 43    & 111   & 286080 & 1.000     & 1.000  & 1.000     & 383.09                  \\ \hline
Random Forest Tree                                      & 735457 & 26927 & 5065   & 281126 & 0.971    & 0.969   & 0.970      & 179.18                 \\ \hline
\rowcolor[HTML]{EFEFEF} 
Gradient Boosted Tree                                   & 762310 & 74    & 4417   & 281774 & 0.996     & 0.996  & 0.996     & 2586.35                 \\ \hline
\end{tabular}
}
\end{table*}

\begin{table*}[!htb]
\caption{Performance measures for Botnet attacks in CSE-CIC-IDS2018 dataset with four active workers}
\label{table4}
\resizebox{1\textwidth}{!}{
\begin{tabular}{|l|c|c|c|c|c|c|c|c|}
\hline
\rowcolor[HTML]{C0C0C0} 
\multicolumn{1}{|c|}{\cellcolor[HTML]{C0C0C0}Algorithm} & TN     & FP    & FN     & TP     & Precision & Recall & F-Measure & Training Time (Seconds) \\ \hline
Logistic Regression                                     &  691043   & 71341  & 148246  &137945 & 0.779     & 0.791  & 0.779     & 1968.61                \\ \hline
\rowcolor[HTML]{EFEFEF} 
Support Vector Machine                                  & 715711 & 46673  & 286189 & 2      & 0.519     & 0.683  & 0.590     & 3310.17                \\ \hline
Bernoulli naive Bayes                                   &  762304  & 80    & 441   & 285750 & 1.000     & 1.000  & 1.000     & 83.53                  \\ \hline
\rowcolor[HTML]{EFEFEF} 
Single Decision Tree (Tuned)                            & 762349  & 35    & 112    & 286079 & 1.000     & 1.000  & 1.000     & 363.54                  \\ \hline
Random Forest Tree                                      &  730803 & 31581 & 7864   & 278327 &   0.964     & 0.962  & 0.963     & 187.16                  \\ \hline
\rowcolor[HTML]{EFEFEF} 
Gradient Boosted Tree                                   &  762306 & 78    & 4417   & 281774 & 0.996     & 0.996  & 0.996     & 2246.44                 \\ \hline
\end{tabular}
}
\end{table*}

\begin{table*}[!htb]
\caption{Performance measures for Botnet attacks in CSE-CIC-IDS2018 dataset with five workers}
\resizebox{1\textwidth}{!}{
\label{table5}
\begin{tabular}{|l|c|c|c|c|c|c|c|c|}
\hline
\rowcolor[HTML]{C0C0C0} 
\multicolumn{1}{|c|}{\cellcolor[HTML]{C0C0C0}Algorithm} & TN     & FP    & FN     & TP     & Precision & Recall & F-Measure & Training Time (Seconds) \\ \hline
Logistic Regression                                     & 691043 & 71341  & 148246  & 137945  & 0.779     & 0.791  & 0.779     & 1910.76                 \\ \hline
\rowcolor[HTML]{EFEFEF} 
Support Vector Machine                                  & 715711 & 46673  & 286189 & 2      & 0.519     & 0.683  & 0.590     & 3304.51               \\ \hline
Bernoulli naive Bayes                                   &  762304  & 80    & 441   & 285750 & 1.000     & 1.000  & 1.000     & 69.49                \\ \hline
\rowcolor[HTML]{EFEFEF} 
Single Decision Tree (Tuned)                            & 762339 & 45    & 112    & 286079 & 1.000     & 1.000  & 1.000     & 360.96                  \\ \hline
Random Forest Tree                                      & 735553 & 26831 & 3905   & 282286 & 0.972    & 0.971  & 0.971     & 184.51                  \\ \hline
\rowcolor[HTML]{EFEFEF} 
Gradient Boosted Tree                                   & 762304 & 80    & 4419   & 281772 & 0.996     & 0.996  & 0.996     & 2231.94                 \\ \hline
\end{tabular}
}
\end{table*}

\begin{table*}[!htb]
\caption{Performance measures for Botnet attacks in CSE-CIC-IDS2018 dataset with six workers}
\resizebox{1\textwidth}{!}{
\label{table6}
\begin{tabular}{|l|c|c|c|c|c|c|c|c|}
\hline
\rowcolor[HTML]{C0C0C0} 
\multicolumn{1}{|c|}{\cellcolor[HTML]{C0C0C0}Algorithm} & TN     & FP    & FN     & TP     & Precision & Recall & F-Measure & Training Time (Seconds) \\ \hline
Logistic Regression                                     &  691043 & 71341 & 148246 & 137945 & 0.779     & 0.791  & 0.779     & 1839.46                  \\ \hline
\rowcolor[HTML]{EFEFEF} 
Support Vector Machine                                  & 715711  & 46673 & 286189 & 2     & 0.519     & 0.683  & 0.590     & 3404.44                \\ \hline
Bernoulli naive Bayes                                   &  762304 & 80    &  441    & 285750 & 1.000     & 1.000  & 1.000     & 66.45                  \\ \hline
\rowcolor[HTML]{EFEFEF} 
Single Decision Tree (Tuned)                            & 762348 & 36    & 114    & 286077  & 1.000     & 1.000  & 1.000     &  392.06               \\ \hline
Random Forest Tree                                      & 730927 & 31457 & 7012   & 279179 & 0.965      & 0.963  & 0.964     & 192.12                 \\ \hline
\rowcolor[HTML]{EFEFEF} 
Gradient Boosted Tree                                   &  762303  & 81    &  4413    & 281778 & 0.996     & 0.996  & 0.996     & 2221.98                 \\ \hline
\end{tabular}
}
\end{table*}

\begin{table*}[!htb]
\caption{Performance measures for Botnet attacks in CSE-CIC-IDS2018 dataset with seven workers}
\label{table7}
\resizebox{1\textwidth}{!}{
\begin{tabular}{|l|c|c|c|c|c|c|c|c|}
\hline
\rowcolor[HTML]{C0C0C0} 
\multicolumn{1}{|c|}{\cellcolor[HTML]{C0C0C0}Algorithm} & TN     & FP    & FN     & TP     & Precision & Recall & F-Measure & Training Time (Seconds) \\ \hline
Logistic Regression                                     &691043 & 71341 & 148246 &  137945 & 0.779     & 0.791  & 0.779     & 1789.41                 \\ \hline
\rowcolor[HTML]{EFEFEF} 
Support Vector Machine                                  & 715711 & 46673  & 286189 & 2      & 0.519     & 0.683  & 0.590     & 3434.41                \\ \hline
Bernoulli naive Bayes                                   &  762304  & 80    & 441   & 285750 & 1.000     & 1.000  & 1.000     & 63.76                  \\ \hline
\rowcolor[HTML]{EFEFEF} 
Single Decision Tree (Tuned)                            & 762342 & 42   & 108   & 286083 & 1.000     & 1.000  & 1.000     & 299.45               \\ \hline
Random Forest Tree                                      & 730137 &  32247 & 6429   & 279762 &  0.965     &   0.963  &  0.964     & 180.65                  \\ \hline
\rowcolor[HTML]{EFEFEF} 
Gradient Boosted Tree                                   & 762304 & 80    & 4419   & 281772 & 0.996     & 0.996  & 0.996     &  2128.01                \\ \hline
\end{tabular}
}
\end{table*}

\begin{table*}[!htb]
\caption{Performance measures for Botnet attacks in CSE-CIC-IDS2018 dataset with eight workers}
\label{table8}
\resizebox{1\textwidth}{!}{
\begin{tabular}{|l|c|c|c|c|c|c|c|c|}
\hline
\rowcolor[HTML]{C0C0C0} 
\multicolumn{1}{|c|}{\cellcolor[HTML]{C0C0C0}Algorithm} & TN     & FP    & FN     & TP     & Precision & Recall & F-Measure & Training Time (Seconds) \\ \hline
Logistic Regression                                     & 691043 & 71341  &148246 & 137945 & 0.779     & 0.791  & 0.779     & 1455.03               \\ \hline
\rowcolor[HTML]{EFEFEF} 
Support Vector Machine                                  & 715711 & 46673 & 286189 & 2    & 0.519     & 0.683  & 0.590     & 3435.17                \\ \hline
Bernoulli naive Bayes                                   & 762304  & 80    & 441   & 285750 & 1.000     & 1.000  & 1.000     & 58.48                  \\ \hline
\rowcolor[HTML]{EFEFEF} 
Single Decision Tree (Tuned)                            & 762344 & 40    & 111    & 286080 & 1.000     & 1.000  & 1.000     & 267.12                 \\ \hline
Random Forest Tree                                      & 740669 & 21715 & 3972   & 282219  &  0.977     &  0.976  & 0.976  &  203.63                  \\ \hline
\rowcolor[HTML]{EFEFEF} 
Gradient Boosted Tree                                   & 762332 & 52   & 4421   &  281770 & 0.996     & 0.996  & 0.996     &  2045.21                 \\ \hline
\end{tabular}
}
\end{table*}

From these tables, in terms of performance measures, no change is observed when increasing the number of worker nodes in Logistic Regression, Support Vector Machine, and Bernoulli naive Bayes algorithms except for training time. In Single Decision Tree and Gradient Boosted Tree algorithms, only a small change can be observed in the number of correctly and incorrectly classified instances, but the overall statistical measures such as Precision, Recall, and F-Measure remains constant. This is due to a small element of randomness in most Decision Tree algorithms when determining which and how many samples to use. This small change could also be observed when running the same experiments multiple times in client mode. The element of randomness in the random forest tree algorithm is higher due to random subspace selection compared to other decision tree algorithms, thus a substantial change can be observed in the number of classified instances as well as performance measures in different cluster setups. {\color{red} The Logistic Regression model's training time was reduced from 4653.09 Seconds in single-node mode to 2436.68 seconds in the three-node cluster setup. This shows the scaling capability of this algorithm for our scenario. Support Vector Machine shows some scaling problems after using more than 3 worker nodes and this the same for Single Decision Tree after using more than 4 worker nodes. Bernoulli naive Bayes also shows good scalability in the performed experiments.}

\begin{figure*}[!h]
\centering
\includegraphics[scale=0.75]{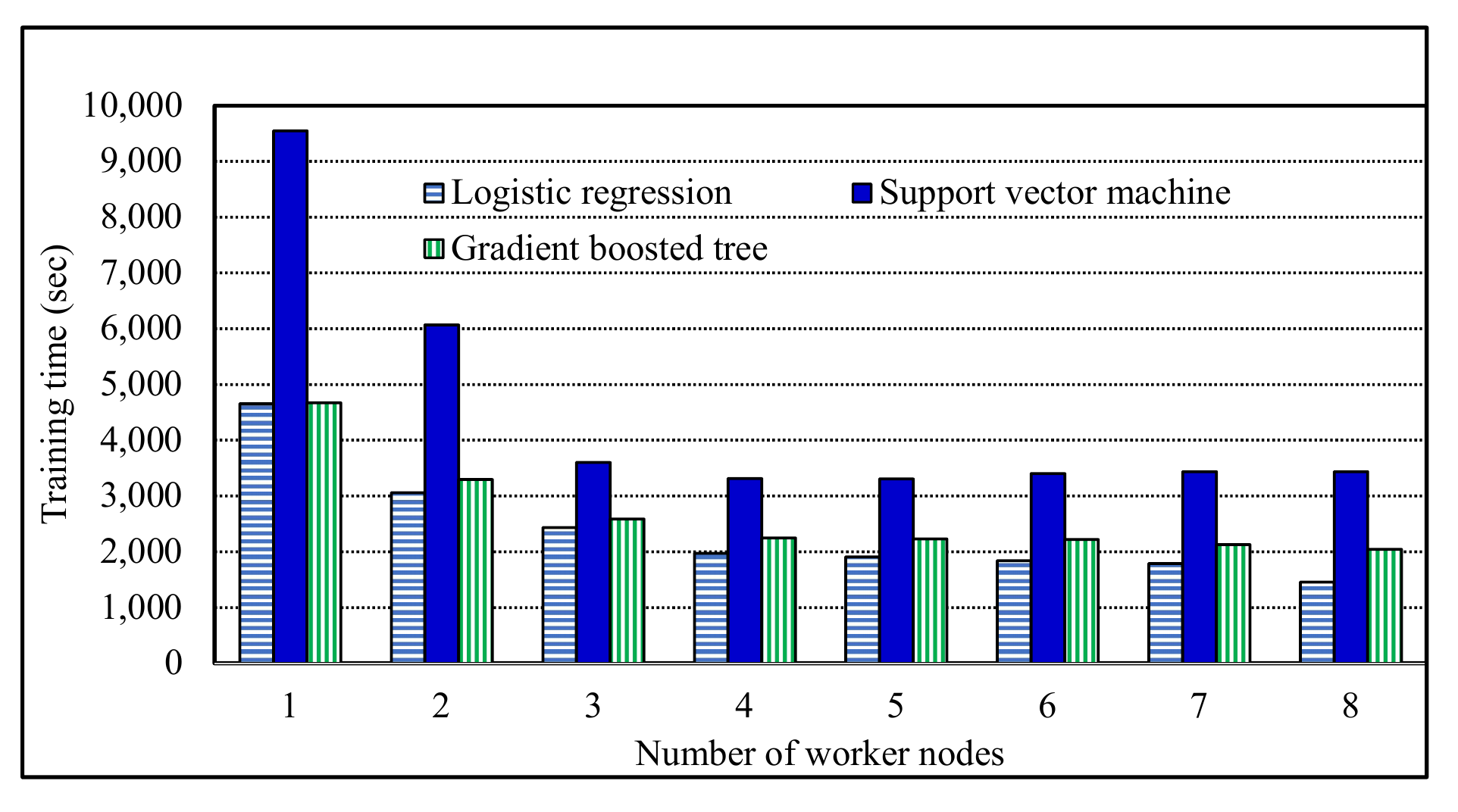}
\caption{The effect of number of nodes on training time in Bernoulli naive Bayes, Single Decision Tree and Random Forest Tree algorithms}
\label{LSG}
\end{figure*}

Scalability results are presented in two separate tables for better readability. Figure \ref{LSG} depicts the scalability of Logistic Regression, Support Vector Machine, and Gradient Boosted Tree algorithms using Spark.
As shown in this Figure, the Logistic Regression algorithm is almost straightforward to be parallelized across a Spark cluster. So the Spark platform can be recommended to process massive data for Logistic Regression problems. For Support Vector Machine, we can observe that more Speed-up is not achieved by using more than five worker nodes. The reason could be many redundant communications needed to send a copy of data to each slave machine in this algorithm. The same outcome occurs for the Gradient Boosted Tree algorithm when using more than four worker nodes.

\begin{figure*}[!h]
\centering
\includegraphics[scale=0.75]{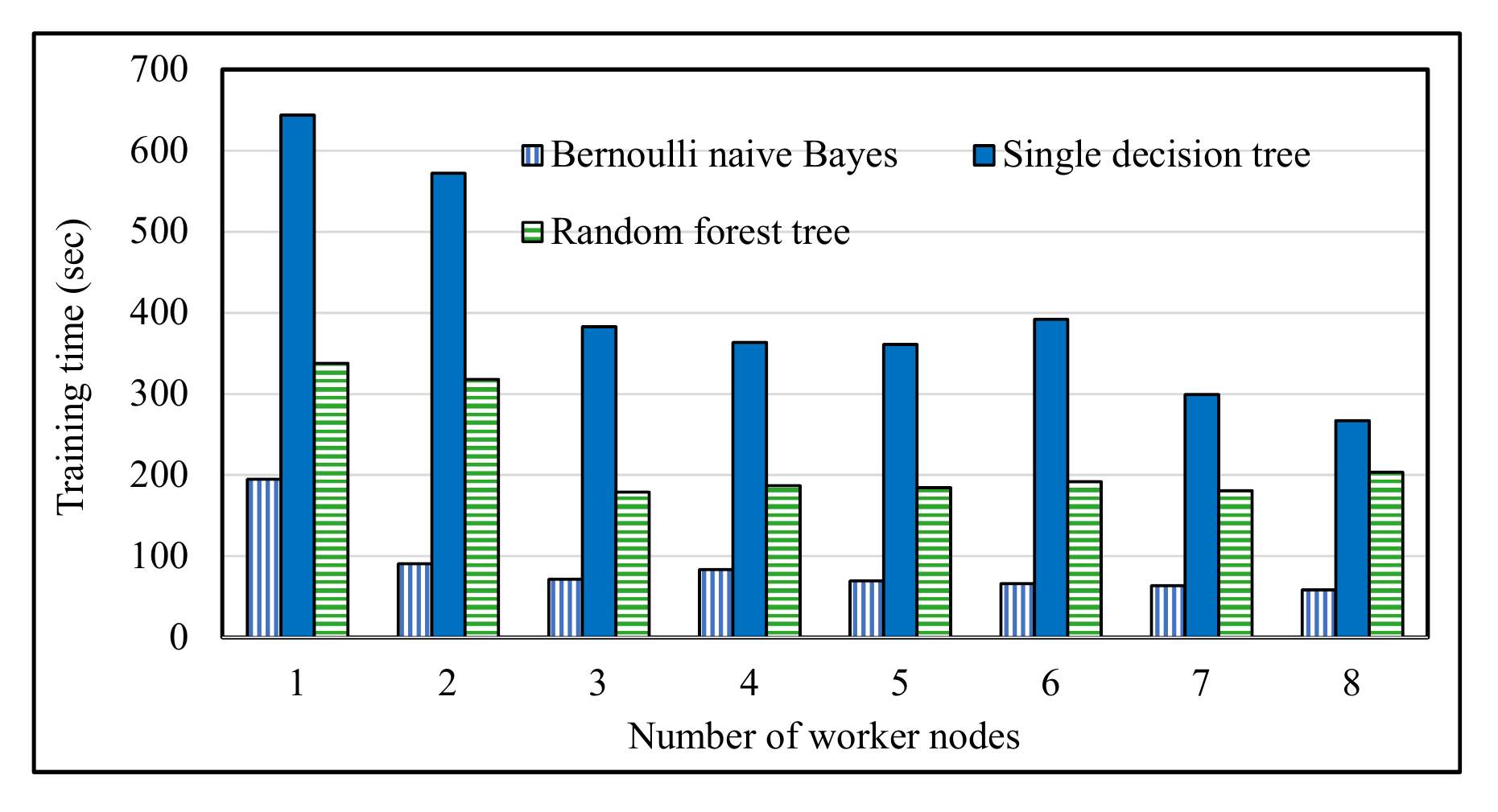}
\caption{The effect of number of nodes on training time in Logistic Regression, Support Vector Machine and Gradient Boosted Tree algorithms}
\label{BSR}
\end{figure*}

\begin{figure*}[!h]
\centering
\includegraphics[width=0.9\textwidth]{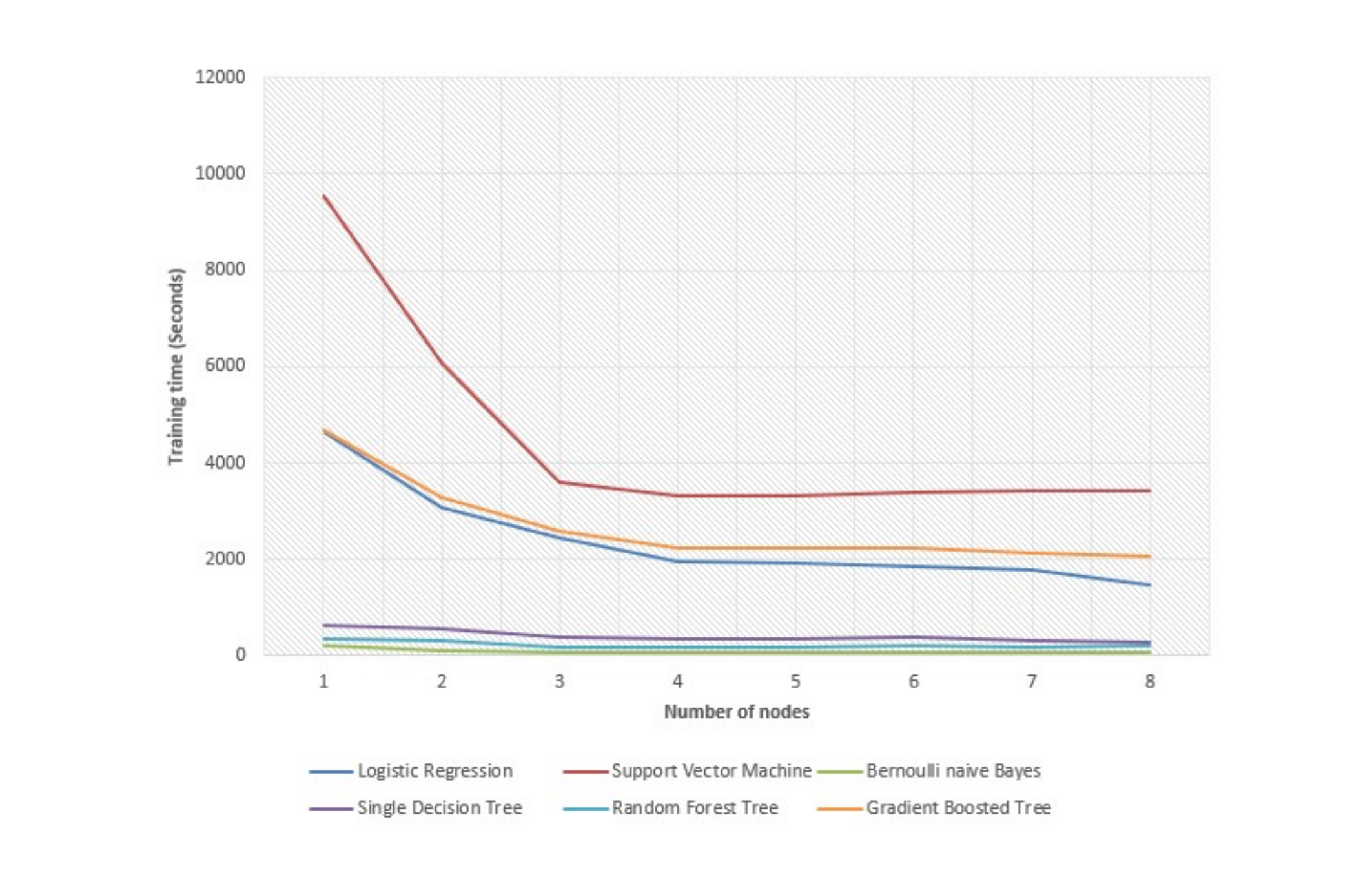}
\caption{{\color{red} The overall effect of number of nodes on training time}}
\label{fig-line}
\end{figure*}

Figure \ref{BSR} depicts the scalability of Bernoulli naive Bayes, Single Decision Tree, and Random Forest Tree algorithms using Spark. As depicted in this Figure, the Naive Bayes model is highly scalable. The probabilities for each record in this algorithm can be calculated independently so the speed-up can be observed by using more worker nodes in our dataset with around one million records. In decision tree algorithms, some random operations are involved, but in general, the Single Decision Tree shows more speed-up compared to the Random Forest Tree algorithm. {\color{red} The overall effect of number of nodes on training time is also depicted in \ref{fig-line}, maximum speed-up threshold for worker nodes can be obtained from this chart. }

In \cite{bekkerman2011scaling}, only 208,904 records are used with 249 incorrectly classified instances while the number of incorrectly classified instances is 157 in the worst case of our experiments by using 1,048,575 instances.
\section{Conclusion}
\label{conclusion}
In this paper, a distributed approach for network intrusion detection using a well-known dataset and six machine learning algorithms has been presented. The goal was to speed up the training phase in processing large amounts of network traffic features and solve the problem of memory management for intrusion detection. We demonstrated that the naive Bayes algorithm could be highly parallelizable by the Apache Spark framework and achieve the highest speed-up due to independent calculations for probabilities. Hyperparameter tuning can help to achieve results with the best accuracy measures. The overall results show that MLlib can be used to train memory-intensive algorithms like SVM and high depth decision trees and also speed up the training time for some applications using more worker nodes in a cluster setup.

\section*{Acknowledgement}
This work has been supported by the Razi University incident response team under grant number 97P101.

\nocite{*} 

\bibliographystyle{unsrtnat}
\bibliography{ref}

\end{document}